\documentclass{elsart}
\usepackage{graphicx}
\begin{document}
\begin{frontmatter}
\journal{Surface Science}
\title{A first principles study of sub-monolayer Ge on Si(001)}
\author{J.~Oviedo\thanksref{birm}}, 
\thanks[birm]{Present address: Nanoscale Physics Research Laboratory,
School of Physics and Astronomy, University of Birmingham,
Birmingham B15 2TT, UK}
\ead{jol@us.es}
\author{D.R.~Bowler\corauthref{drb}} \and
\ead{david.bowler@ucl.ac.uk}
\ead[url]{http://www.cmmp.ucl.ac.uk/$\sim$drb/research.html}
\corauth[drb]{Corresponding author}
\author{M.J.~Gillan}
\ead{m.gillan@ucl.ac.uk}
\address{Department of Physics and Astronomy, University College London,
Gower Street, London WC1E 6BT, UK}

\begin{abstract}
Experimental observations of heteroepitaxial growth of Ge on Si~(001)
show a $( 2 \times n )$ reconstruction for sub-monolayer coverages,
with dimer rows crossed by missing-dimer trenches.  We present
first-principles density-functional calculations designed to elucidate
the energetics and relaxed geometries associated with this
reconstruction. We also address the problem of how the formation
energies of reconstructions having different stoichiometries should be
compared. The calculations reveal a strong dependence of the formation
energy of the missing-dimer trenches on spacing $n$, and demonstrate
that this dependence stems almost entirely from elastic
relaxation. The results provide a natural explanation for the
experimentally observed spacings in the region of $n \simeq 8$.
\end{abstract}

\end{frontmatter}

\section{\label{sec:intro}Introduction}

Understanding the heteroepitaxial growth of Ge on Si(001) is vitally
important for two reasons: first, it is a prototypical system for
strained, Stranski-Krastanow growth; second, it has great potential
for growing new semiconductor devices while remaining compatible with
existing Group IV technology.  We present here a first-principles
investigation of the energetics of the $( 2 \times n )$ reconstruction
observed at low Ge coverage, with the aim of explaining why the
periodicity $n$ has the values observed experimentally; we also
present an interpretation of the energetics in terms of the relaxed
atomic geometries.

\begin{figure}
\includegraphics[clip,width=\columnwidth]{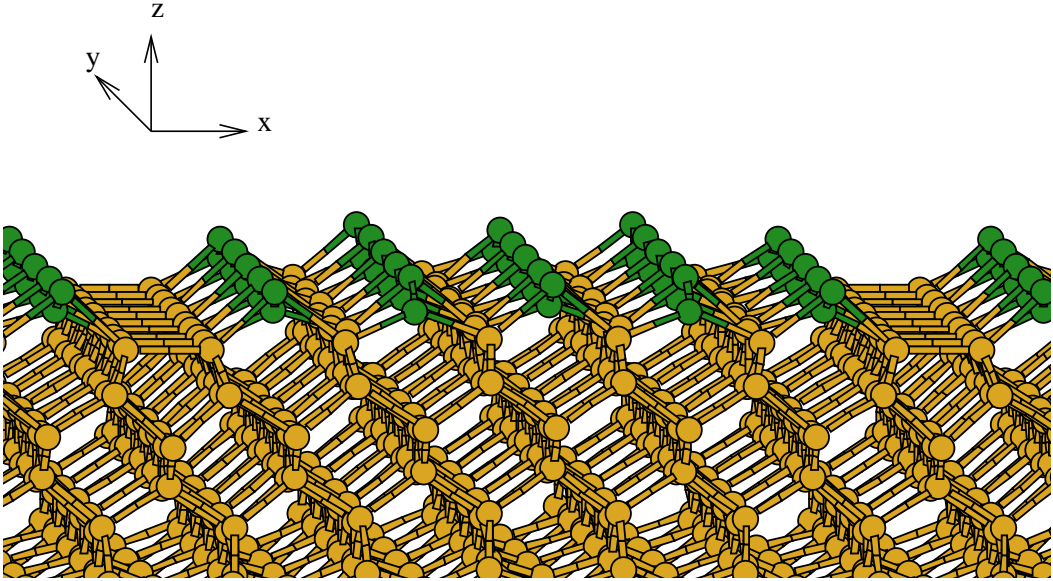}
\caption{\label{fig:Trenches}Geometry of the $(2 \times n)$ 
reconstruction of Ge on Si(001) for the case $n=6$, with light and
dark coloured spheres representing Si and Ge.  Dimer rows, 
missing-dimer trenches and normal to the surface lie along
mutually perpendicular x, y and z directions.}
\end{figure}

The clean Si~(001) surface shows the well-known reconstruction due to
the formation of rows of Si dimers. During the early stages of
solid-source molecular beam epitaxy (SSMBE) of Ge on Si~(001), the
growth is remarkably similar to that of Si itself, with rows of dimers
forming on the surface. As the coverage approaches one monolayer,
however, the system shows the effects of strain due to the mismatch of
the Si and Ge lattice parameters, and trenches of missing dimers
appear (see Fig.~\ref{fig:Trenches}). These trenches are
oriented at right angles to the axis of the dimer rows, and there is a
fairly regular spacing $n$ between
trenches~\cite{Kohler1992,Tersoff1991,Tersoff1992,Chen1994,Liu1996,Voigt1999}.
This is the $(2 \times n )$ reconstruction, with $n$ observed to be
about~8, though the distribution of $n$ is fairly broad and somewhat
dependent on growth conditions, with values up to~12 being reported. A
similar reconstruction is observed during gas-source MBE
(GSMBE)~\cite{Goldfarb1997a,Goldfarb1997b,Huang1997}, though there are
small differences due to the presence of hydrogen on the surface.

As further Ge is deposited (both in SSMBE and GSMBE), a series of
further reconstructions are seen, which vary depending on growth
conditions.  The temperature at which growth occurs, the deposition
rate of Ge and the presence of hydrogen all affect the
growth~\cite{Goldfarb1997a}.  However, the next stage observed after
the formation of $(2 \times n )$ is generally the $ ( m \times n )$
reconstruction~\cite{Kohler1992,Goldfarb1997a,Goldfarb1997b}.  This
forms when a further layer of Ge grows on a $( 2 \times n )$ surface,
and does not fill in the trenches in the surface.  The increased
strain in the new layer generally means that the new periodicity ($m$)
is less than the old ($n$).  On Si~(001), there are two step types:
parallel to the dimer rows (A-type, generally smooth) and
perpendicular to the dimer rows (B-type, generally
rough)~\cite{Chadi1987,Bowler1998}.  After one or two monolayers of Ge
have been deposited, this roughness is seen to first equalise and then
reverse~\cite{Goldfarb1997a,Wu1995}.  Finally, large scale features
form along the elastically soft (100) and (010) directions: ``hut''
pits and
clusters~\cite{Kohler1992,Goldfarb1997a,Goldfarb1997b,Mo1991}.
Although the work reported here focuses only on the $( 2 \times n )$
reconstruction, it provides the foundation for investigating the other
more complicated reconstructions. In particular, we will report
elsewhere~\cite{Li2002} on tight-binding investigations of the $( m
\times n )$ reconstruction, which use the present first-principles
calculations to validate a tight-binding parameterisation of the Si/Ge
system for this kind of application.

The calculations to be presented are based on density-functional
theory (DFT) in the generalized-gradient approximation (GGA),
implemented with pseudopotentials and plane-wave basis sets.  Starting
from the perfect Ge monolayer as point of reference, we have
calculated the formation energy and relaxed structure of missing-dimer
trenches having spacings $n$ ranging from 4 to 12. By calculating the
trench formation energy both with and without atomic relaxation, we
shall show that the strong dependence of formation energy on $n$
arises almost entirely from relaxation effects. We shall also show
that, in a sense that is appropriate to the usual experimental
conditions, this formation energy is a minimum for a value of $n$ in
the experimentally observed range.

There has been previous modelling of the Si/Ge~$( 2 \times n )$
reconstruction, but only using empirical potentials, such as the
Stillinger-Weber and modified Keating
forms~\cite{Tersoff1991,Tersoff1992,Liu1996}.  These also studied
atomic relaxation and strain effects, relating these to the
periodicity of the reconstruction, and found results in broad
agreement with experiment. Our calculations provide support for the
physical mechanisms that emerged from these empirical studies. One
problem addressed in the earlier work, but in our opinion not fully
resolved, was that of comparing the energies of surface structures
having different stoichiometries. This problem necessarily arises if
one wishes to compare the energies of $( 2 \times n )$ reconstructions
having different $n$ values, since removal or addition of Ge is needed
to go from one to the other. The solution to this problem is
equivalent to assigning an appropriate chemical potential to Ge, and
different ways of doing this have been proposed. Since the predicted
equilibrium value of $n$ depends on the choice of chemical potential,
it is essential to identify the choice that corresponds to the real
experimental conditions. We shall outline here what we believe to be
the correct procedure. We note that during the experimental growth
process there may be some intermixing of Ge and
Si~\cite{Voigt1999,Huang1997}, though this can be suppressed by
surfactants such as As or H.  The likely effect is a reduction of
surface strain, and hence an increase in the value of
$n$\cite{Voigt1999}; for the growth of the first monolayer, such
effects are expected to be small, and we neglect them in the present
work.

Technical details of our calculations are summarised in the next
Section. We then present (Sec.~\ref{sec:energetics}) our results for
energetics and relaxed geometry, first of the perfect Ge monolayer,
and then of the missing-dimer trenches, including our analysis of the
trench formation energy into electronic and relaxation contributions.
Sec.~\ref{sec:expt} outlines our arguments about the correct Ge
chemical potential to use in determining the equilibrium inter-trench
spacing, and gives our numerical result for this, which is close to
the experimental value $n \simeq 8$.  The paper ends with a summary of
our conclusions.

\section{\label{sec:details}Computational Methods}

The fundamental ideas of DFT~\cite{Hoh1964,Kohn1965} have been
extensively reviewed (see e.g. Refs.~\cite{Jones1989,Payne1992}), as
have the pseudopotential and plane-wave
techniques~\cite{Payne1992}. The present calculations were performed
using the VASP code~\cite{Kresse1996}, and employ the standard
ultra-soft pseudopotentials~\cite{Vanderbilt1990} that form part of
the code. The approximation we use for exchange-correlation energy is
the generalised-gradient approximation (GGA) due to Perdew and Wang
(PW91)~\cite{Wang1991,Perdew1992}.  The choice of GGA rather than the
local-density approximation (LDA) is deliberate. Since the energetics
that interests us here depends quite sensitively on bonding and
rebonding effects, and since the errors in bond energies are generally
much larger with LDA than with GGA (LDA generally overbinds
significantly), we regard the use of GGA as essential in this work.

DFT/pseudopotential/plane-wave calculations are most easily performed
in periodic boundary conditions, and we therefore adopt the periodic
slab geo\-metry usually employed for surface-science work. The
scientific issues we are addressing require the accurate treatment of
quite small energy differences (typically on the order of 100~meV),
and we have made efforts to ensure that the calculations are fully
converged with respect to the thickness of the slabs and the width of
the vacuum layer separating neighbouring slabs.  For an eight layer
$(2 \times 1)$ reconstructed slab, we found that the total energy was
converged to better than 1 meV for a vacuum layer of 5 \AA\ (compared
to 8 \AA).  For this vacuum width, we found that the change in surface
energy in going from an eight layer slab to a twelve layer slab was
less than 1 meV per dimer.  (Both tests were conducted with a
plane-wave cutoff energy $E_{\rm cut}$ of 225 eV and a $4 \times 4
\times 1$ Monkhorst-Pack \textbf{k}-point mesh\cite{Monkhorst1976}.)
Given the results of these tests, we chose to perform all the main
Si/Ge calculations with eight-layer slabs, the top layer being
Ge. Both top and bottom surfaces of the slab are reconstructed to form
dimers, as further specified below. The width of the vacuum gap was
taken to be 5~\AA. Careful attention to basis-set completeness and
Brillouin-zone sampling is also essential. In general, $E_{\rm cut}$
was chosen to be 225~eV, and $k$-point sampling was performed using
the $4 \times 4 \times 1$ Monkhorst-Pack mesh, but detailed evidence
will be presented below about the convergence of our results with
respect to these parameters.

\section{\label{sec:results}Results}

\subsection{\label{sec:energetics}Energetics and geometries}

\begin{figure}
\includegraphics[clip,width=\columnwidth]{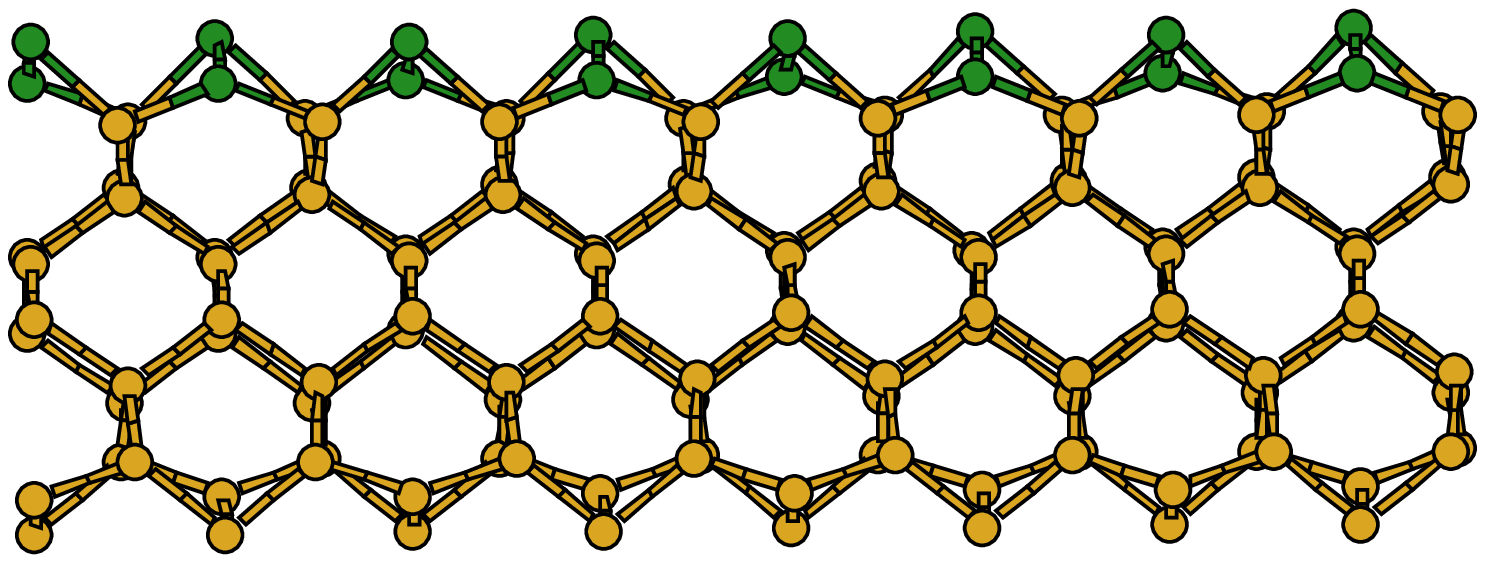}
\includegraphics[clip,width=\columnwidth]{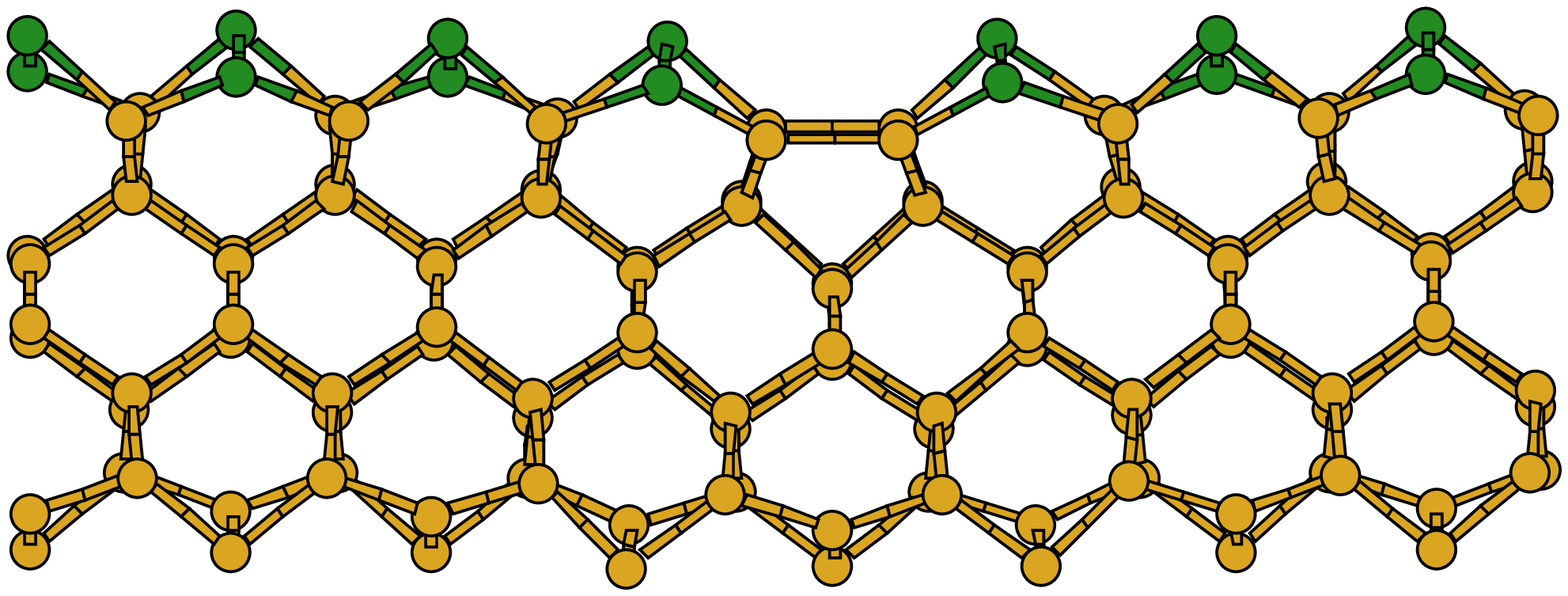}
\caption{\label{fig:GeML}Calculated structures for (a) (top) a perfect
Ge monolayer and (b) (bottom) a relaxed Ge layer with the spacing
$n=8$ between missing-dimer trenches.  In both structures, darker
atoms (top surface) are Ge and all other atoms are Si.}
\end{figure}

We begin by summarising the structure and energetics of the system
having a perfect Ge monolayer in the $p ( 2 \times 2 )$ reconstruction
(Fig.~\ref{fig:GeML}(a)). For the fully relaxed system, we find the
following structural parameters (results from
Refs.~\cite{Cho1994a,Cho1994b,Kruger1994,Tang1994,Jenkins1996} in
parentheses): Ge-Ge bond length 2.55~\AA\ ($2.38-2.44$~\AA); Ge-dimer
tilt angle 19.2$^\circ$ ($14.2-18.5^\circ$); up and down Ge-Si bond
lengths 2.48 and 2.39~\AA\ (2.42 and 2.34~\AA).  We note that the
methods used to obtain the earlier results compared with here differ
from ours in two significant ways: first, they used LDA rather than
GGA; second, they used the $( 2 \times 1 )$ rather than the $p ( 2
\times 2 )$ reconstruction. Both of these factors may have an
appreciable effect on the structure of the monolayer; we note
particularly that use of the $( 2 \times 1 )$ reconstruction may well
prevent the favourable relaxations along the dimer row allowed by $p (
2 \times 2 )$. Given these differences, we regard the agreement with
previous results as reasonable.

For comparison, we note the corresponding parameters for the $p ( 2
\times 2 )$ reconstruction on the clean Si~(001) surface, obtained
using the same GGA, and with the GGA lattice parameter for Si ($a_0 =
5.45$~\AA, corresponding to a bulk bond length of 2.36~\AA): dimer
Si-Si bond length 2.36~\AA; dimer-second layer up and down bond
lengths 2.40 and 2.34~\AA.  The Ge-Ge dimer bond is elongated
(actually beyond the bulk bond length, which we found to be 2.49~\AA),
with the extra freedom to relax allowing the substrate to take on more
bulk-like lengths and angles.

We also report here the Ge monolayer formation energy per Ge dimer,
denoted by $E_{\rm m}$, since it will be needed later.  We define this
to be the energy change per Ge dimer when we start with the perfect
relaxed Si~(001)~$2 \times 2$ surface and bring isolated Ge atoms from
infinity to form the perfect relaxed $2 \times 2$ Ge monolayer. We
find the value $E_{\rm m} = - 9.64$~eV, and we have checked that this
is converged within 0.01~eV with respect to slab thickness, plane-wave
cut-off and $k$-point sampling.

We now turn to the energetics of formation of missing-dimer
trenches. The results come from a series of calculations in which the
periodically repeated cell contains a single missing-dimer trench,
with spacings $n = 4$, 6, 8, 10 and 12 between trenches. In all cases,
the repeating cell is orthorhombic. Referring to
Fig.~\ref{fig:Trenches} and denoting the lengths of the three cell
edges by $X$ (along the dimer rows), $Y$ (along the missing-dimer
trenches) and $Z$ (normal to the surface), these lengths are: $X = ( n
/ \sqrt{2}) a_0$, $Y = \sqrt{2} a_0$, and $Z = 15.88$~\AA, where $a_0
= 5.44$~\AA. The value of $Z$ is appropriate to the eight-layer slab
with vacuum width of 5.0~\AA\ used in all the calculations (see
Sec.~\ref{sec:details}).

We first report values for the fully relaxed missing-dimer formation
energy $E_{\rm f} ( n )$, defined as the energy per removed Ge-dimer
needed to form an array of missing-dimer trenches with spacing $n$. In
this process, we start from Si~(001) with a non-defective monolayer of
Ge, and the removed Ge atoms are taken to infinity, where they are
isolated atoms in free space.  In computational terms, we express
$E_{\rm f} ( n )$ as the difference $E_{\rm fin} - E_{\rm init}$ of
the fully relaxed energies per repeating cell of two systems, both
having the same orthorhombic cell of dimensions $(X,Y,Z)$. The initial
system (energy $E_{\rm init}$) is the slab with a single non-defective
monolayer of Ge on one face. The final system (energy $E_{\rm fin}$)
is formed from the initial system by removing one Ge-dimer from each
repeating cell to create an array of infinite missing-dimer trenches with
spacing $n$. We use exactly the same orthorhombic cells for the two
systems, with the same plane-wave cut-off and $k$-point sampling,
because this aids cancellation of errors.

We report in Table~\ref{tab:convErm} our values for $E_{\rm f} ( n )$
for the series of $n$ values.  All the results were obtained for the
eight-layer slab with the vacuum gap of 5 \AA\ chosen for the reasons
explained in Sec.~\ref{sec:details}.  To show that our results are
converged with respect to basis-set completeness and $k$-point
sampling, we report values of $E_{\rm f} ( n )$ for a moderate (150
eV) and a large (225 eV) plane-wave cut-off energy $E_{\rm cut}$ and
for two $k$-point sets ($4\times 4\times 1$ and $8\times 8\times
1$). It is clear that $E_{\rm f} ( n )$ is converged to within a few
meV with respect to $k$-point sampling and plane-wave cut-off when we
use a $4\times 4\times 1$ \textbf{k}-point mesh and a plane-wave
cut-off of 225 eV.  We note that $E_{\rm f} ( n )$ is a monotonically
decreasing function of $n$, which attains a plateau value for $n >
8$. Its overall variation with $n$ is very substantial, since it is
nearly 1~eV higher for $n = 4$ than for $n = 12$. This effective
repulsion between missing-dimer trenches for spacings below $n \sim 8$
has been found before in calculations based on empirical
models~\cite{Tersoff1991,Tersoff1992,Liu1996}. Its consequence is that
for a given overall density of trenches they will tend to become
equally spaced.

\begin{table}
\caption{\label{tab:convErm}Calculated values (eV units) of the fully
relaxed missing-dimer trench formation energy $E_{\rm f}(n)$
as a function of inter-trench spacing $n$ (see text for detailed 
definition).  Results are given for different values of plane-wave
cut-off energy $E_{\rm cut}$ and for different \textbf{k}-point 
sampling meshes.}
\begin{tabular}{cccc} 
\hline
$n$&$E_{\rm cut}$=150 eV, $4 \times 4 \times 1$   &  $E_{\rm cut}$=225 eV, 
$4 \times 4 \times 1$ & $E_{\rm cut}$=225 eV, $8 \times 8 \times 1$    \\  
\hline
4 &    9.819   &  9.882  & 9.881   \\ 
6 &    9.378   &  9.449  & 9.448   \\ 
8 &    9.179   &  9.232  & 9.234   \\ 
10 &   9.150   &  9.206  & 9.208   \\ 
12 &   9.081   &  9.182  & 9.185   \\
\hline
\end{tabular}     
\end{table}

\begin{figure}
\includegraphics[clip,width=\columnwidth]{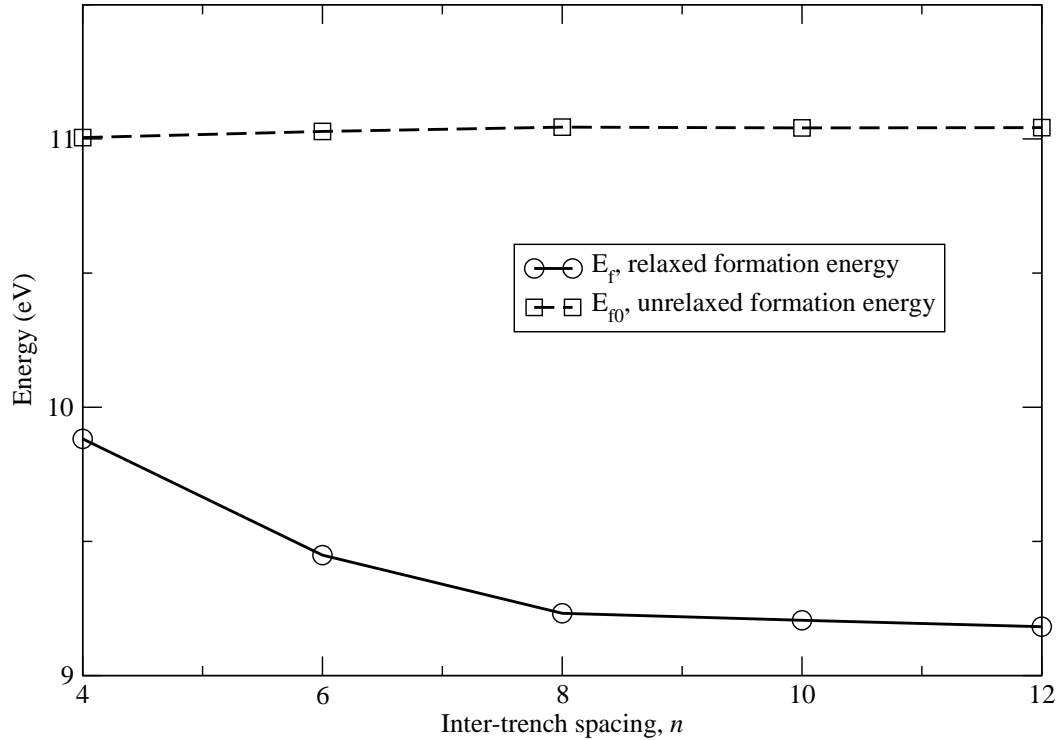}
\caption{\label{fig:ErelaxN}Calculated values of relaxed (circles,
solid line) and unrelaxed (squares, dashed line) formation energies
$E_{\rm f}$ and $E_{\rm f}^0$ of missing-dimer trenches in a Ge
monolayer on Si(001).}
\end{figure}

Previous work~\cite{Tersoff1991,Tersoff1992,Liu1996} suggests that the
effective repulsion between trenches stems from the elastic relaxation
field surrounding each missing dimer. To test this, we have repeated
the calculations of $E_{\rm f} ( n )$, but without relaxation (we
denote the unrelaxed value by $E_{\rm f}^0 ( n )$).  As before, we
start from the fully relaxed initial system, but when the Ge atoms are
removed, all atoms are held fixed in their initial positions. The
resulting $E_{\rm f}^0 ( n )$ values are reported as a function of $n$
in Fig.~\ref{fig:ErelaxN}, and we see that their variation with $n$ is
extremely small. This means that all the variation in the fully
relaxed $E_{\rm f} ( n )$ values comes from the relaxation of the
final system with respect to the initial system. The relaxation energy
$E_{\rm f} - E_{\rm f}^0$ has a magnitude of nearly 2~eV for widely
spaced missing dimers.

We show our calculated relaxed structure of the trench system in
Fig.~\ref{fig:GeML}(b) for the spacing $n = 8$. The key feature to
notice is the large inward relaxation towards the trench, leading to
Si--Si rebonding across the trench. Quantitatively, the relaxed Si--Si
distance across the trench is 2.54~\AA, to be compared with the bond
length of 2.36~\AA\ in the Si perfect crystal at ambient pressure.
This is an elongation of $\sim8$\%, showing the strain that the system
is under.  It is interesting to compare this result with the single
missing dimer in Si(001), which has an identical structure, apart from
the Ge in the top layer.  There, the bonds across the trench have a
length of 2.56~\AA~\cite{Wang1993,Owen1995}; the similarity suggests
that the limiting factor on the relaxation in both cases is the Si-Si
distance.  The Ge dimers neighbouring the trench show an inward
relaxation of 0.65~\AA\ from their positions in the perfect
monolayer. The up and down Si--Ge bond lengths neighbouring the trench
are 2.58 and 2.45~\AA, so that there is a significant lengthening
compared with the perfect Ge monolayer.  As expected, for smaller
values of $n$, the relaxation around the trench is less
pronounced. For example, in the $n = 4$ case, the Si--Si separation
across the trench is 2.63~\AA, and the inward relaxation of the Ge
dimers neighbouring the trench is 0.48~\AA. The suppression of
relaxation for small spacings is clearly responsible for the strong
$n$-dependence of $E_{\rm f} ( n )$, corresponding to the effective
repulsion between trenches.

\subsection{The experimental missing-dimer trench spacing}
\label{sec:expt}

We now want to use our results for $E_{\rm f} ( n )$ to clarify why
the missing-dimer trenches adopt the spacings $n$ in the range $8 -
12$ observed in experiments.  We explain first the
statistical-mechanical basis for the arguments to be used. We assume
that the surface is in thermal equilibrium, so that, for a given
number of Ge atoms on the surface, the probability of finding any
particular arrangement $\gamma$ of these atoms is proportional to
$\exp ( - E_\gamma / k_{\rm B} T )$, where $E_\gamma$ is the energy of
$\gamma$. (Strictly speaking, $E_\gamma$ should be a
non-configurational {\em free} energy, but here we take it to be the
equilibrium energy of $\gamma$.) Our thermal equilibrium assumption
means that we are ignoring kinetic effects. We will discuss the
validity of this assumption in Sec.~\ref{sec:conc}.

Since our approach is to discuss the arrangements that will be seen in
thermal equilibrium for a {\em given} number of Ge atoms on the
surface, and we need only know how $E_\gamma$ varies as we go from one
arrangement of these atoms to another, the energy zero chosen for
$E_\gamma$ is irrelevant. However, it will be convenient to relate
$E_\gamma$ values to the energy of one particular arrangement of Ge,
which we call the `reference' arrangement, whose energy is $E_{\rm
ref}$. We choose this to be the arrangement in which all Ge atoms form
perfect dimers, which are arranged to make a non-defective monolayer
covering a certain area of the surface. The shape of the boundary of
this area does not make any difference, but the following arguments
become simpler if it is rectangular.

We now consider the energies of arrangements created from the
reference arrangement by the formation of missing-dimer trenches. For
definiteness, let there be $L$ dimer rows each containing $P$ dimers
in the reference arrangement. We form $Q$ equally spaced missing-dimer
trenches by removing $Q L$ dimers, replacing them at the boundary of
the monolayer, and allowing the whole system to relax. It is
convenient to divide this process into two parts: (i) the prior
fetching of $2 Q L$ Ge atoms from infinity and their deposition at the
boundary of the reference system, in such a way that each dimer row is
increased in length from $P$ dimers to $P + Q$ dimers, the number of
dimer rows remaining the same; (ii) the subsequent formation of the
missing-dimer trenches by the removal of $Q L$ Ge dimers and their
separation to infinity. We write the energy in process (i) as $Q L
E_{\rm p}$, where $E_{\rm p}$ is a constant energy that will be
discussed further below.  The energy change in process (ii) is $Q L
E_{\rm f} ( n )$, where $E_{\rm f} (n)$ is the trench formation energy
defined above, with the spacing $n$ given by $n = ( P + Q ) / Q$. The
total energy change $\Delta E$ is therefore:
\begin{equation}
\Delta E = Q L ( E_{\rm f} ( n ) + E_{\rm p} ) = 
\frac{L P}{n - 1} ( E_{\rm f} ( n ) + E_{\rm p} )  =
L P \zeta ( n ) \; ,
\label{eq:zeta}
\end{equation}
where $\zeta ( n ) = ( E_{\rm f} ( n ) + E_{\rm p} ) / ( n - 1 )$.
The energetically most favorable value of $n$ is therefore obtained by
minimizing $\zeta ( n )$ with respect to $n$

\begin{figure}
\includegraphics[clip,width=\columnwidth]{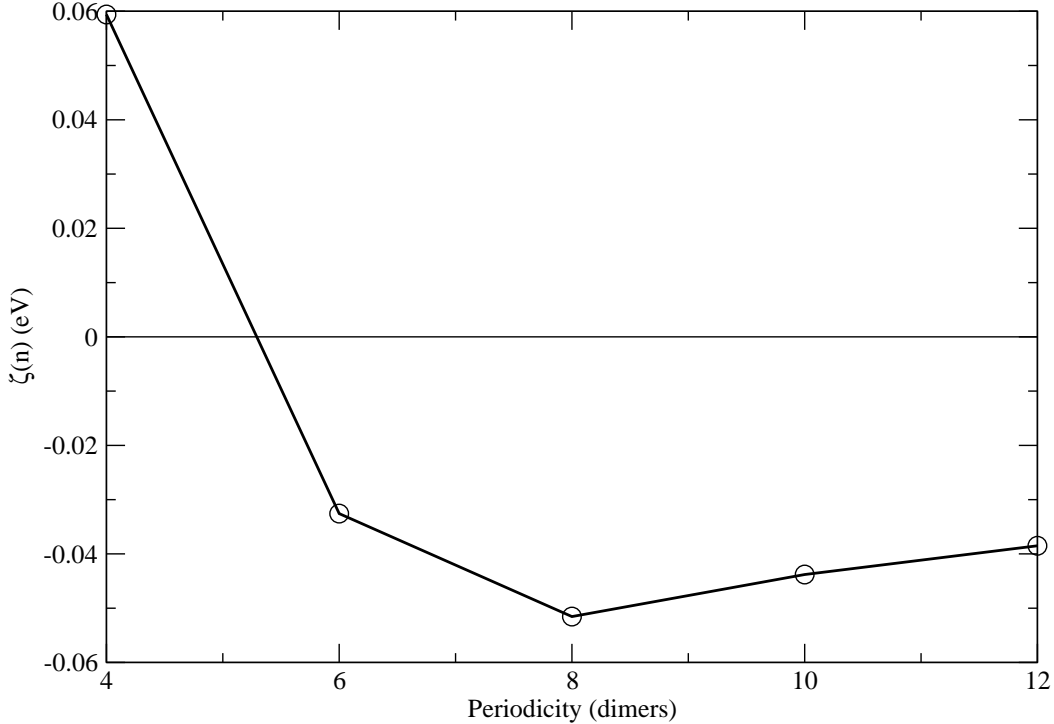}
\caption{\label{fig:zeta}Calculated values (eV units) of $\zeta(n)
\equiv (E_{\rm f}(n) + E_{\rm m}) / (n-1)$ as a function of
inter-trench spacing $n$, with $E_{\rm f}(n)$ the formation energy of
missing-dimer defect trenches and $E_{\rm m}$ the Ge monolayer
formation energy (see Eq.~\protect\ref{eq:zeta} and text).}
\end{figure}

In fact, the energy $E_{\rm p}$ of process (i) is the same as the
monolayer formation energy $E_{\rm m}$ already discussed in
Sec.~\ref{sec:energetics}. The reason is that the energy change on
forming a certain amount of monolayer cannot depend on whether this is
accomplished by placing Ge atoms on the clean Si surface, or by
bringing Ge atoms to the boundary of a pre-existing piece of
monolayer. (This assumes, of course, that we ignore edge effects,
which is an appropriate approximation here.)

We now use our calculated $E_{\rm m} = - 9.64$~eV/dimer (see above,
Sec.~\ref{sec:energetics}) to obtain numerical values for $\zeta ( n
)$, which we present in Fig.~\ref{fig:zeta}. We note that $\zeta ( n
)$ is negative for large $n$, so that it is energetically favourable to
form widely spaced missing dimers. However, the repulsion between rows
causes $\zeta ( n )$ to increase at small $n$, and it has a minimum at
$n \simeq 8$.  This optimum $n$ value corresponds well to the typical
spacing observed experimentally.

\section{\label{sec:conc}Discussion and conclusions}

A number of key points have emerged from our first-principles
calculations on the $( 2 \times n )$ reconstruction of sub-monolayer
Ge on Si~(001). First, we have shown that the non-defective Ge
monolayer is energetically unstable with respect to formation of
widely spaced ($n \rightarrow \infty$) missing-dimer trenches. Second,
there is a substantial effective repulsion between trenches, so that
their formation becomes energetically unfavourable for small $n$. This
means that in thermal equilibrium there is an optimal value of $n$,
for which our calculations yield the estimate $n \simeq 8$, in
respectable agreement with the typical values observed
experimentally. Third, we have shown that the effective repulsion is
entirely due to elastic relaxation effects: the energy lowering due to
relaxation is greatest when the strain fields of different trenches do
not overlap, and decreases as $n$ decreases. Fourth, we have seen that
for large $n$ there is significant rebonding between Si atoms in the
trench, and this is probably essential in making trench formation
energetically favourable.

The picture we have established is not entirely new. The work based on
empirical models\cite{Tersoff1991,Tersoff1992,Liu1996} referred to in
Sec.~\ref{sec:intro} came to essentially the same conclusion about the
important role of elastic effects. The more fundamentally based
first-principles calculations presented here therefore provide support
for the earlier models. However, we have emphasised that a correct
identification of the appropriate Ge chemical potential is crucial in
understanding the energetic stabilisation that results from trench
formation, as well as the equilibrium value of $n$. In earlier work,
it was suggested that this chemical potential should be identified
with the energy of bulk unstrained Ge (for large, fully relaxed
islands), or else Ge biaxially strained to the Si lattice constant
(for wide, coherent islands)~\cite{Tersoff1991}. We have argued here
that the formation energy of the perfect Ge monolayer provides the
experimentally relevant point of reference in fixing this chemical
potential.

Our reasoning is based on the assumption of full thermal equilibrium,
and we have not attempted to account for kinetic effects.  Since we
are only trying to investigate the observed spacing $n$ of the
reconstruction, this is reasonable.  Although a variety of kinetic
effects are observed during the formation of the $(2 \times n)$
reconstruction (including the ``displacive incorporation'' growth
model)~\cite{Tromp1993}, once the complete layer is formed it is
stable, and kinetic effects are unlikely to play a role.  We can also
consider the limit of slow growth conditions, where thermal
equilibrium will be a valid assumption~\cite{Tersoff1991}.  The
observed range of $n$ is rather broad (roughly from 8 to 12), in part
depending upon growth conditions; this is perfectly consistent with
our results, which show only a weak increase in formation energy
beyond the spacing $n=8$ at which it is a minimum.  We also expect the
temperature at which the growth occurs and the growth source to have
an effect.  We have also deliberately ignored intermixing between Ge
and Si layers.  There is MEIS evidence showing that for 1ML coverage
(i.e. equivalent to the $(2 \times n)$ surface we are modelling) at
low temperatures (up to 500$^\circ$C) there is little
intermixing~\cite{Tromp1993,Copel1990}, though recent measurements and
calculations~\cite{Uberuaga2000} indicate intermixing starting at about
500$^\circ$C.  Nevertheless, the simplest model is one without
intermixing, and this is where we have started.  Intermixing will
reduce the surface strain, thus increasing the value of $n$, as seen
in experiment.  Our value could thus be considered a lower limit,
taken for thermal equilibrium and segregated Si and Ge layers.

\ack 
JO and DRB thank EPSRC for funding through grants M01753 and M71640.
DRB also thanks the Royal Society for funding through a University
Research Fellowship.  Calculations were performed on the Manchester
CSAR service, through an allocation of time to the UKCP consortium
(EPSRC grant M01753) and at the HiPerSPACE Centre at UCL (JREI grant
JR98UCGI).

\bibliography{Defects}

\end{document}